\documentclass[12pt]{article}
\usepackage{graphicx}

\begin{document}
\title{GLOBAL COMPARISONS OF MEDIANS AND OTHER QUANTILES IN A ONE-WAY DESIGN WHEN THERE ARE TIED VALUES}
\author{Rand R. Wilcox \\
Dept of Psychology \\
University of Southern California\\
}
\maketitle
\pagebreak
\begin{center}
ABSTRACT
\end{center}

For $J \ge 2$ independent groups, the paper deals with testing the global 
 hypothesis that all $J$ groups have a common
population median or identical quantiles, with an emphasis on the quartiles.  
 Classic rank-based methods are sometimes
suggested for comparing medians, but it is well known that under general conditions
they do not adequately address this goal. 
Extant methods  based on the usual sample median are unsatisfactory 
when there are tied values except for the special case $J=2$.
 A variation of the percentile bootstrap used in conjunction with the Harrell--Davis quantile estimator performs well in simulations.
 The method is illustrated with data  from the Well Elderly 2 study.



\section{Introduction}

For $J$ independent random variables, 
let $\theta_j$ be the population median or 
some other quantile
associated with $j$th variable ($j=1, \ldots, J$).
The paper considers the problem of testing 
\begin{equation}
H_0: \theta_1= \cdots = \theta_J,  \label{null}
\end{equation}
particularly when there are tied values. The focus is on comparing the medians as well as
the upper or lower quartiles, but it is evident that the results are relevant when comparing other
quantiles instead.

For the special case $J=2$, the Wilcoxon-Mann-Whitney  (WMW) test is sometimes suggested for  comparing medians, but it is well known that under general conditions
it does not accomplish this goal  (e.g., Hettmansperger, 1984; Fung, 1980).  Roughly, the reason is that
for two  independent random variables, $X$ and $Y$, 
it is not based on an estimate of $\theta_1-\theta_2$, but rather on an estimate of  $P(X<Y)$.
Another concern is that when distributions differ, under general conditions the WMW test
 uses the wrong standard error (e.g., Cliff, 1996; Wilcox, 2012). 
More generally, the Kruskall--Wallis test, which reduces to the  Wilcoxon-Mann-Whitney test when $J=2$, does not test (1).

Yet another possible approach is to use the usual sample median in conjunction with a permutation test. However, results in Romano (1990)
establish that this approach is unsatisfactory as well.  

For a random sample $X_1, \ldots, X_n$, let $X_{(1)} \leq \ldots  \leq X_{(n)}$ denote the observations written in
ascending order and let $M_j$ denote the usual sample median.
That is, 
if the number of observations, $n$, is odd,
\[M = X_{(m)},\]
where $m=(n+1)/2$ and 
if  $n$ is even,
\[M = \frac{(X_{(m)}+X_{(m+1)})}{2},\]
where now $m=n/2$.
A  natural way of proceeding is to estimate $\theta_j$ with  $M_j$ and use some test statistic that is
based in part on some estimate of  the standard error of  $M_j$. 
When sampling from a continuous distribution where tied values occur with
probability zero, an effective method was studied by Bonett and  Price  (2002)
 that can be used when $J=2$ or when $J>2$ and the goal is to test
some hypothesis based on a linear 
contrast of the population medians. Numerous methods for estimating the 
standard error of $M_j$ have been derived, but extant results 
 indicate  that all of them can perform poorly when tied values can 
 occur (Wilcox, 2012). Wilcox  (2006) found a slight extension of a standard 
percentile bootstrap method that performs well when testing (1),
 there are tied values and $J=2$. But Wilcox (2012) notes that in terms of
testing (1) when $J>2$,  evidently no method has been found that performs 
well in simulations when there are tied values. 

There is another complication when working with the usual sample median. It is well known 
that when sampling from a continuous distribution,
under certain regularity conditions,  $M_j$ is asymptotically normal. 
However, when sampling from a discrete distribution with a finite sample space,
$M_j$ does not converge to a normal distribution.   More broadly, when estimating quantiles using a single order statistic, or a weighted average of  two order statistics,
assuming asymptotic normality is generally unsatisfactory when dealing with discrete distributions where tied values occur. 

As an illustration, consider a beta-binomial distribution having  
the probability function 
\begin{equation}
P(x) = \frac{\mathbf{B}(m-x+r,x+s)}
{(m+1)\mathbf{B}(m-x+1,x+1)\mathbf{B}(r,s)}, \label{bbdist}
\end{equation}
where $\mathbf{B}$ is the complete beta function, $r>0$ and $s>0$ are parameters that determine the shape of the distribution and 
$x=0, \ldots, m$.  Consider $m=30$. Then the cardinality of the sample space
is 31 and as is evident, if the sample size is $n>31$, tied values occur with
probability one.
The left panel of Figure 1 shows a plot of 3000 sample medians generated from a beta-binomial distribution with $r=1$ and $s=3$ based on a sample size 
 $n=100$. (So the  beta-binomial distribution  is skewed to the right,  $P(x)$ is monotonic decreasing 
and $x=6$ corresponds to the .52 quantile.)
The right panel is the same as the left, only now $n=500$. As is evident, the sampling distribution has not moved closer to a normal
distribution and indeed the cardinality of the sample space has decreased, indicating that
any method for making inferences based on the sample median that assumes asymptotic normality can perform poorly. 






 \begin{figure}
\resizebox{\textwidth}{!}
{\includegraphics*[angle=0]{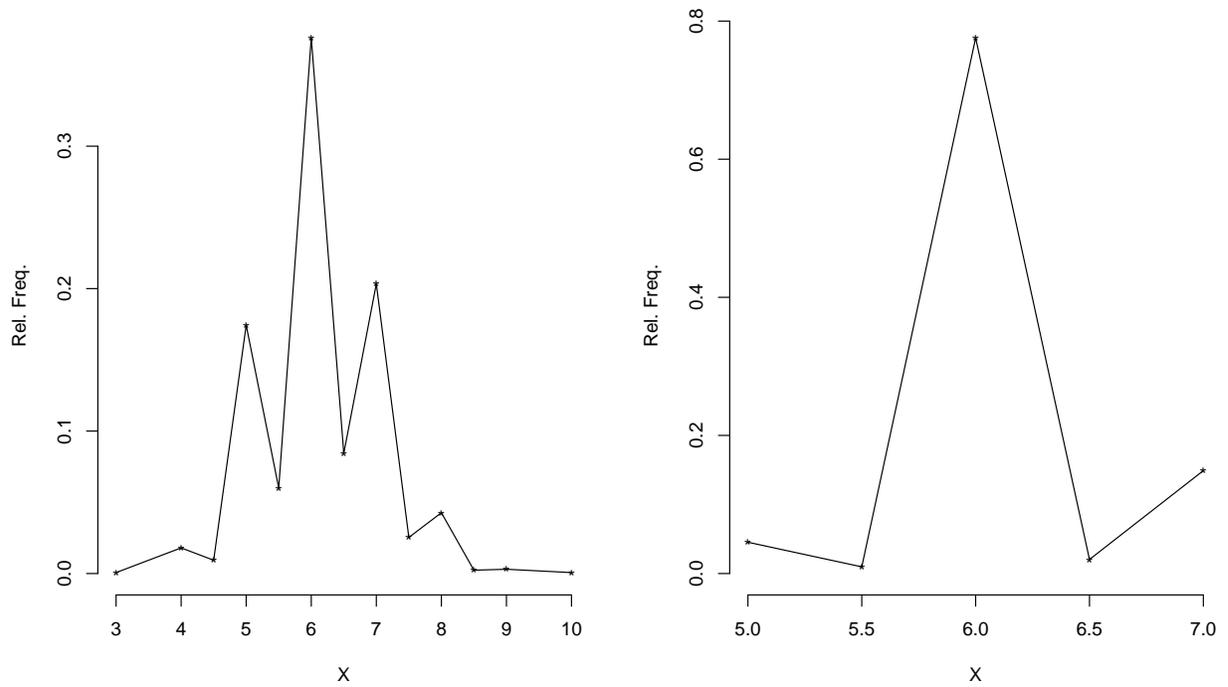}}
\caption{The left panel shows a plot of 3000 sample medians when sampling from a beta-binomial distribution, $n=100$, $m=30$, $r=1$, $s=3$. The right panel is 
a plot of the 3000 medians when $n=500$}
\end{figure}

For the special case where the goal is to compare two independent groups, 
a method for comparing quantiles that deals effectively with tied values is to use 
a percentile bootstrap in conjunction with the quantile estimator derived
by Harrell and Davis (1982); see Wilcox et al. (2013). 
The Harrell--Davis estimator uses a weighted average of all the order statistics. The 
result is a sampling distribution that is typically well approximated by a continuous distribution. 
Consider again the right panel of Figure 1 and note that
the cardinality of the sample space is only five. 
That is, only five values for the sample median were observed among the 3000 estimates.
In contrast, if the Harrell--Davis estimator is used, there are no tied values among all 3000 estimates. 
But the method studied by Wilcox et al. is limited
to $J=2$;  there are no results on how best to proceed when testing (1) and there are $J>2$ independent groups.

Here, two methods for testing (1) were considered, both of which were based on the Harrell--Davis estimator.  
The first was based on a test statistic mentioned by Schrader and
Hettmansperger (1980), and studied by
He, Simpson  and Portnoy (1990). The basic strategy was to use a percentile bootstrap method to estimate the null distribution. But situations were found
where this approach performed very poorly when dealing with tied values, so further details are omitted. 
The other method is described in section 2 and 
simulation results are reported in section 3. 
Section 4 illustrates the proposed method using data from the Well Elderly 2 study.
The strategy is not new and has been found to perform reasonably well when dealing with
 other robust measures of location (Wilcox, 2012). 
 However, when using the usual sample median for the situation at hand, preliminary
 simulations found that it performs poorly in terms of controlling the Type I error probability. 
 Results in  Wilcox et al. (2013) suggest that using
 instead the Harrell--Davis estimator, reasonably good control over the Type I error probability might be obtained.  
 So the goal  here is to determine 
 the extent to which this is the case. 

Note that in terms of characterizing the typical value of some random variable, the population median is an obvious choice. However, situations
are encountered where differences occur in the tails of a distribution that have substantive interest (e.g. Doksum \& Sievers, 1976; Wilcox et al., 2014). 
This issue can be addressed by comparing quantiles other than the median, which can help provide a deeper understanding of how distributions
differ. This point is illustrated in section 4. 

 \section{Description of the Proposed Method}

To describe the Harrell and Davis (1982)  estimate of the $q$th quantile,
 let $Y$ be a random variable having a beta distribution with parameters
$a=(n+1)q$ and $b=(n+1)(1-q)$. That is, the probability density function of $Y$ is
\[\frac{\Gamma(a+b)}{\Gamma(a)\Gamma(b)}y^{a-1}(1-y)^{b-1}, \, 0 \le y \le 1,\]
where $\Gamma$ is the gamma function. Let
\[W_i = P\left(\frac{i-1}{n} \le Y \le \frac{i}{n}\right).\]
For a random sample $X_1, \ldots, X_n$, let $X_{(1)} \leq \ldots  \leq X_{(n)}$ denote the observations written in
ascending order.
The Harrell--Davis estimate of $\theta_q$, the $q$th quantile, is
\begin{equation}
\hat{\theta}_{q} = \sum_{i=1}^n W_i X_{(i)}. \label{hdest}
\end{equation}
In terms of its standard error, Sfakianakis and Verginis (2006) show that in some situations the Harrell--Davis estimator competes well with alternative estimators that again use a weighted average of all the order statistics, but there are exceptions.
(Sfakianakis and Verginis derived alternative estimators that have advantages over the Harrell--Davis in some situations. But 
when sampling from heavy-tailed distributions, the standard errors of their estimators can be substantially larger than the standard error of $\hat{\theta}_{q}$.) Additional comparisons of various estimators
are reported by Parrish (1990),
Sheather and Marron (1990),
as well as Dielman, Lowry and Pfaffenberger (1994). 
 The only certainty is that no single estimator dominates in terms
 of efficiency. For example, the Harrell--Davis estimator has a smaller standard error than
  the usual sample median when sampling from a normal
 distribution or a distribution that has relatively light tails, but for sufficiently heavy-tailed distributions,
  the reverse is true (Wilcox, 2012, p. 87).

Let $\delta_{jk}=\theta_j-\theta_k$. Roughly, the strategy for testing (1) is to test
\begin{equation}
H_0: \delta_{jk}=0,  \, \,  \forall \,   j<k,   \label{altnull}
\end{equation} 
via a percentile bootstrap method. 
If the null hypothesis is true, then $\bf{0}$, a vector of zeros having length $(J^2-J)/2$, should be nested fairly deeply within a 
cloud of bootstrap estimates of $\delta_{jk}$. Moreover, the depth of $\bf{0}$ can be used to compute a p-value as will be indicated.
A  natural way of measuring the depth of $\bf{0}$ within a bootstrap cloud is via Mahalanobis distance.
 Note, however,  that the $\delta_{jk}$ parameters are linearly dependent. This indicates that 
  Mahalanobis distance can fail from a computational point of view because the bootstrap  covariance matrix will be singular. 
 This proved to be the case, so the strategy here is to measure the depth of  $\bf{0}$ using a method that does not
 require the use of a covariance matrix. 
  
 For completeness, note that the issue of a singular covariance matrix could be avoided by using the first group as
 a reference group and testing $H_0$: $\theta_1-\theta_2 = \theta_1-\theta_3 = \cdots  = \theta_1-\theta_J$.
  In terms of a Type I error, this approach is reasonable, but in terms of power, this is not necessarily the case. 
  The reason is that 
   power can depend on which group is used as the reference group because the choice of the reference group impacts
   the magnitude of the differences between the medians that are compared. 

 To describe the details of the proposed test of (1) via (4), 
 let $X_{ij}$ be a random sample from the $j$th group ($i=1, \ldots, n_j$).
 Generate a bootstrap sample from the $j$th group by resampling with replacement $n_j$ 
 observations from group $j$. Let
 $\hat{\theta}_j^*$ be the estimate of the $q$th quantile for group $j$ based on this bootstrap sample. Let $\hat{\delta}_{jk}^*=\hat{\theta}_j^*-\hat{\theta}_k^*$,
 $j<k$. Repeat this process $B$ times
 yielding $\hat{\delta}^*_{bjk}$, $b=1, \ldots, B$. 
 Here, $B=600$ is used in order to avoid overly high execution time and because this choice has been found to provide reasonably good control over the
 Type I error probability when dealing with related bootstrap techniques (e.g., Wilcox, 2012). However, in terms of power, there might be a practical advantage 
 to using a larger choice for $B$ (Racine \& MacKinnon,  2007).

A portion of the strategy used  here is based on measuring the depth of a point in a multivariate data cloud using 
a projection-type method, which provides an approximation of half-space depth (Wilcox, 2012, section 6.2.5).
For notational convenience, momentarily focus on an $n \times p$ matrix of data, ${\bf Y}$. Let $\hat{\tau}$ be some measure of location based on 
${\bf Y}$. For simplicity, the marginal medians (based on the usual sample median) are used. 
 Let
\[{\bf U}_i = {\bf Y}_i - \hat{\tau}\]
($i=1, \ldots, n$), 
\[
C_i=  {\bf U}_i {\bf U}_i^{\prime}
\]
and for any $j$ ($j=1, \ldots, n$),  let
\[W_{ij} = \sum_{k=1}^J U_{ik} U_{jk},\]
\begin{equation}
T_{ij} = \frac{W_{ij}}{C_i} (U_{i1}, \ldots, U_{ip}) \label{bigt}
\end{equation}
and
\[D_{ij} =  \| T_{ij}\|,\]
where $\| T_{ij}\|$ is the Euclidean norm associated with the
vector $T_{ij}$ ($i=1,\ldots n$; $j=1, \ldots, n$).
Let 
\[d_{ij} = \frac{D_{ij}}{q_{i2}-q_{i1}},\]
where $q_{i2}$ and $q_{i1}$ are estimates of the upper and lower quartiles, respectively,  based on 
$D_{i1}, \ldots, D_{in}$. Here, $q_{i2}$ and $q_{i1}$ are estimated with  the so-called ideal fourths (e.g., Friqqe et al., 1989.), which
are computed as follows. 
Let  $j$ be the integer portion of $(n/4)+(5/12)$ and  let
\[h=\frac{n}{4} +\frac{5}{12}- j.\]
The lower quartile is estimated with s
\begin{equation}
q_{i1} = (1-h)D_{i(j)} + hD_{i(j+1)},
\end{equation}
where $D_{i(1)} \le \cdots \le D_{i(n)}$.
Letting $k=n-j+1$,  the upper quartile is
\begin{equation}
q_{i2} = (1-h)D_{i(k)} + hD_{i(k-1)}.
\end{equation}

The projection distance of ${\bf Y}_j$,  relative to the cloud of points represented by  ${\bf Y}$,
 is the maximum value of $d_{ij}$, the maximum being taken over $i=1, \ldots, n$.
 This measure of depth is nearly the same as the measure derived by   Donoho and Gasko (1992).
 the only difference is that they used the median absolute difference (mad) as a measure of scale
 rather than the interquartiles range.  MAD has a higher breakdown point but using
 the interquartile range has been found to perform better in various situations 
 (Wilcox, 2012). This might be due to the poor efficiency of mad, but the extent this is the case is 
 unclear.  Perhaps using mad would perform well in the simulations reported here, but
 this is left to future investigations.

Now create a $(B+1) \times L$ matrix ${\bf G}$ where the first $B$ rows are based on the $\hat{\delta}^*_{bjk}$, $b=1, \ldots, B$, $L=(J^2-J)/2$.
That is, row $b$ consists of the $L$ values associated with $\hat{\delta}^*_{bjk}$ for all $j<k$. Row $B+1$ of ${\bf G}$ is a  vector ${\bf 0}$ having length
$L$. Then from general theoretical results in Liu and Singh (1997), 
a (generalized) p-value can be computed based on the relative distance of ${\bf 0}$. Compute the projection distance for each row of ${\bf G}$.
The distance associated with the $b$th row is denoted by $K_b$ and the distance for the null vector (row $B+1$) is denoted by $K_0$.
Then a generalized p-value is 
\[1- \frac{1}{B} \sum_{b=1}^{B} I(K_0 \ge K_b),\]
where the indicator function $I(K_0 \ge K_b)=1$ if $K_0 \ge K_b$, otherwise $I(K_0 \ge K_b)=0$.
This will be called method Q.

\section{Simulation Results}

Simulations were used to study the small-sample
properties of method  Q when there are $J=4$ groups.  Results are reported when comparing 
medians as well as the lower and upper quartiles. 
Estimated Type I error probabilities, $\hat{\alpha}$, were based on 4000 replications.
Both continuous and discrete distributions were used. The
four continuous distributions were
normal, symmetric and heavy-tailed, asymmetric and light-tailed,
and asymmetric and heavy-tailed.
More precisely, four g-and-h distributions were used
(Hoaglin, 1985) that contain the standard  normal distribution as a special case.
If $Z$ has a standard normal distribution, then
\[W = \left\{ \begin{array}{ll}
 \frac{{\rm exp}(gZ)-1}{g} {\rm exp}(hZ^2/2), & \mbox{if $g>0$}\\
  Z{\rm exp}(hZ^2/2), & \mbox{if $g=0$}
   \end{array} \right. \]
has a g-and-h distribution where $g$ and $h$ are parameters that
determine the first four moments.
The four distributions used here were the standard normal ($g=h=0.0$), a
symmetric heavy-tailed distribution ($h=0.2$, $g=0.0$), an asymmetric
distribution with
relatively light tails ($h=0.0$, $g=0.2$), and an asymmetric distribution with
heavy tails ($g=h=0.2$).
Table 1 shows the skewness ($\kappa_1$) and kurtosis
($\kappa_2$)
for each distribution. Additional properties of the g-and-h distribution
are summarized by Hoaglin (1985).

\begin{table}
\caption{Some properties of the g-and-h distribution.}
\centering
\begin{tabular}{ccrr} \hline
g & h &  $\kappa_1$ & $\kappa_2$\\ \hline
0.0 & 0.0  & 0.00 & 3.0\\
0.0 & 0.2 & 0.00 & 21.46\\
0.2 & 0.0  & 0.61 & 3.68\\
0.2 & 0.2 & 2.81 & 155.98\\ \hline
\end{tabular}
\end{table}

As for situations where tied values can occur, consider a discrete distribution
 with a sample space having cardinality $N$. 
A goal in the simulations was to get some sense about how  well method Q controls  the Type I error probability when $N$ is small.
Roughly, as the likelihood of tied values increases, at what point does method Q break down?
Here, results are reported when data are generated from a beta-binomial distribution for which
the cardinality of the sample space is  
 $N=m+1=11$ and $N=m+1=21$. The choices for $(r, s)$ were (3, 3), which has a symmetric distribution, as well  as (1, 3) and (1,9), which are skewed distributions.


First consider  the
four g-and-h distributions when testing at the .05
level and the groups have a common sample size $n=20$. As indicated in Table 2, the estimated Type I error probability ranges between .025 and .062. 
Although the importance of a Type I error depends on the situation, Bradley (1978) suggests that as a general guide, when testing at the .05 level,
the actual level should not drop below .025 or exceed .075. In Table 2, the estimates were in this range.

\begin{table}
\center
\caption{Estimated Type I Error Probability using method Q, continuous distributions, $\alpha=.05$}
\begin{tabular}{ cccccc}
$q$  & $g$ & $h$ & $n=20$  & $n=50$ \\ \hline
0.25 & 0.0 & 0.0 & 0.058 & 0.059\\     
0.25     &    0.0 & 0.2 & 0.031 &   0.046\\ 
0.25     &     0.2 & 0.0 & 0.061 & 0.058\\  
0.25     &    0.2 & 0.2 & 0.038 & 0.055\\ 
0.50 &    0.0 & 0.0 & 0.059 & 0.061\\  
0.50 &   0.0 & 0.2 & 0.046 & 0.057\\  
0.50 &   0.2 & 0.0 & 0.062 & 0.062\\  
0.50 &    0.2 & 0.2 & 0.054 & 0.056\\  
0.75     &     0.2 & 0.0 & 0.049 & 0.054\\ 
0.75     &    0.2 & 0.2 & 0.025 & 0.038\\ \hline   
\end{tabular}
\end{table}


A possible criticism of the results in Table 2 is that they are based on only 4000 replications. Consequently,
some comments about the precision of the estimates in Table 2 are provided. 
Assuming Bradley's criterion is reasonable, consider the issue of whether 
 the actual level is less than or equal .075. 
Using the method in Pratt (1968), it can be seen that based on a two-sided .95 confidence
interval for the actual  level,  the confidence interval will not contain .075
if $\hat{\alpha} \le  .06675$. 
 All of the results in Table 2 suggest that the actual level does not exceed .075. 
 Using instead a .99
confidence interval for the actual level,  $\hat{\alpha} \le .06425$ indicates that 
the actual level does not exceed .075. 
In a similar manner, 
based on a two-sided .95 confidence interval, the confidence interval for the actual level does not contain .025 
if $\hat{\alpha} \ge  .03025$. 
For a .99 confidence interval, $\hat{\alpha} \ge  .032$ is required and there are only two situations
where the estimate is less than .032 which occurred for $n=20$ 
when comparing the quartiles. 


For normal distributions, a simulation was run with $n=100$ as an additional check on how the method
performs as $n$ gets large. The estimated Type I error probability was .058.

For the beta-binomial distributions, estimated Type I error probabilities are shown in Table 3. 
For $m=20$, control over the Type I error probability is reasonably good when comparing medians. But for $m=10$,
it is evident  that  control over the probability of a Type can be unsatisfactory, 
particularly when $(r, s) = (1, 9)$. 
The fact that method Q does not perform well for this 
particular distribution  is not surprising because the .47 quantile is zero. 
When comparing the quartiles with $m=20$, method Q can be unsatisfactory with $n=20$, the highest
estimate of actual level being .085. For this particular situation, 
increasing the sample size of two of the groups to 40, the estimate
 is .064. With all sample sizes equal to 30 and $m=10$, the
estimate is .065. 

Precise details regarding the  rate of convergence to the nominal level is impossible with only
4000 replications, but it is evident that the rate of convergence can depend  on the
nature of the distribution. Among the discrete distributions considered here for which the cardinality of the
sample space is 21, $n \ge 20$ suffices in
terms of achieving an estimated Type I error probability reasonably close to a nominal .05 level when
comparing the medians.
But for $m=10$ (the cardinality of the sample space is 11)
and $(r, s) = (1, 9)$, $n \ge 180$ is  required when comparing medians. 
With $n=100$, for example, the estimate of the actual level exceeds .09, in which case the
.95 confidence interval for the actual level does not contain .075. 


\begin{table}
\caption{Estimated probability of a Type I error using method Q when sampling from a beta-binomial distribution
for a sample sizes $n=20$ and 50, $\alpha=.05$,
and where the cardinality of the sample space is $N=m+1$.}
\centering
\begin{tabular}{ccc cccc}

$q$ & $r$ & $s$ & $(n,m)=(20,10)$ & $(n,m)=(20,20)$ & $(n,m)=(50,10)$  & $(n,m)=(50,20)$ \\ \hline

0.25 & 3 & 3  & 0.074 & 0.071  & 0.068 & 0.063 \\    
0.50 & 3 & 3  &  0.066 & 0.070 &  0.061 & 0.058\\   
0.25 & 1 & 3  &  0.056 & 0.052 & 0.094 & 0.048\\
0.50 & 1 & 3  & 0.059 & 0.062 & 0.060 & 0.060\\
0.75 &   1 & 3  & 0.078 & 0.085 & 0.065 & 0.067 \\    
0.25  &  1 & 9 &  0.008 & 0.058 & 0.000 & 0.067\\   
0.50 &   1 & 9 &   0.088 & 0.052 & 0.154 & 0.050\\  
0.75 &  1 & 9 &  0.061& 0.064 & 0.069 & 0.054\\   
\hline

\end{tabular}
\end{table}

A few simulations were performed using a discrete distribution where the null hypothesis is true but not all of the distributions are identical. All indications are that when
comparing medians, again
the Type I error probability is controlled reasonably well when $n=m=20$. Consider, for example, a beta-binomial distribution where $r=s=3$. Then 11 is the .54 quantile.
Now, suppose that for  all four groups data are generated from a discrete distribution such that $P(X \le x)$ corresponds to a  beta-binomial distribution where $r=s=3$ provided
that  $x \le 11$, but that  otherwise some of the cumulative distributions differ. So the distributions are not all identical in the right tail, but 
 the hypothesis of equal population medians is true. Consider  in particular  the 
situation where
for three of the groups the probability function, say $f(x)$, corresponds to a beta-binomial probability function when $x < 15$. Otherwise
\[f(x)=\sum_{x=15}^{21}  P(x)/7\]
when $x \ge 15$, where again $P(x)$ indicates a   a beta-binomial distribution  given by (\ref{bbdist}). Now the estimated Type I error probability when comparing the population medians is .057.  If instead for $x=15, 16, \ldots, 21$ $f(x)$ is taken to be $P(21)$, $P(20), \ldots, P(15)$, respectively,  the estimated Type I error probability  is .048.
However, when comparing the lower quartiles, now control over the Type I error probability exceeds .09.  Increasing the sample sizes to 40,
this problem persisted. With all sample sizes equal to 50,  control over the Type I error probability is reasonably good, the estimate being .064. 
 





\section{An Illustration}


%






Method Q is illustrated using data from the Well Elderly 2 study (Jackson, et al., 2009; Clark et al. 1997).
Generally, the study dealt with the efficacy of a particular intervention strategy  aimed
at improving the physical and emotional health of older adults. 
One particular issue was whether four educational groups differed in terms of a measure of meaningful activity prior to intervention. 
The four groups were high school graduate, some college or technical school, 4 years of college and  post-graduate school. 
For convenience, these groups are designated G1, G2, G3 and G4 henceforth.
Meaningful activity  was measured with  the sum of 29 Likert items, where the possible responses for each item were 0, 1, 2, 3 and 4.
 Higher scores  reflect higher levels of meaningful  
activities. The sample sizes were 62,  81, 110 and 125, respectively. 

Figure 2 shows boxplots for each of the four groups, 
which suggests that more pronounced differences occur based on the lower quartile compared to 
upper quartile. 
Applying method Q, the p-values corresponding to the .25, .5 and .75 quantiles were 0, .074 and .294, respectively. 
So in terms of participants who
score relatively high on meaningful activities, no significant difference is found, but a significant result is found for 
low levels of meaningful activity as reflected by the .25 quantiles. 


 \begin{figure}
\resizebox{\textwidth}{!}
{\includegraphics*[angle=0]{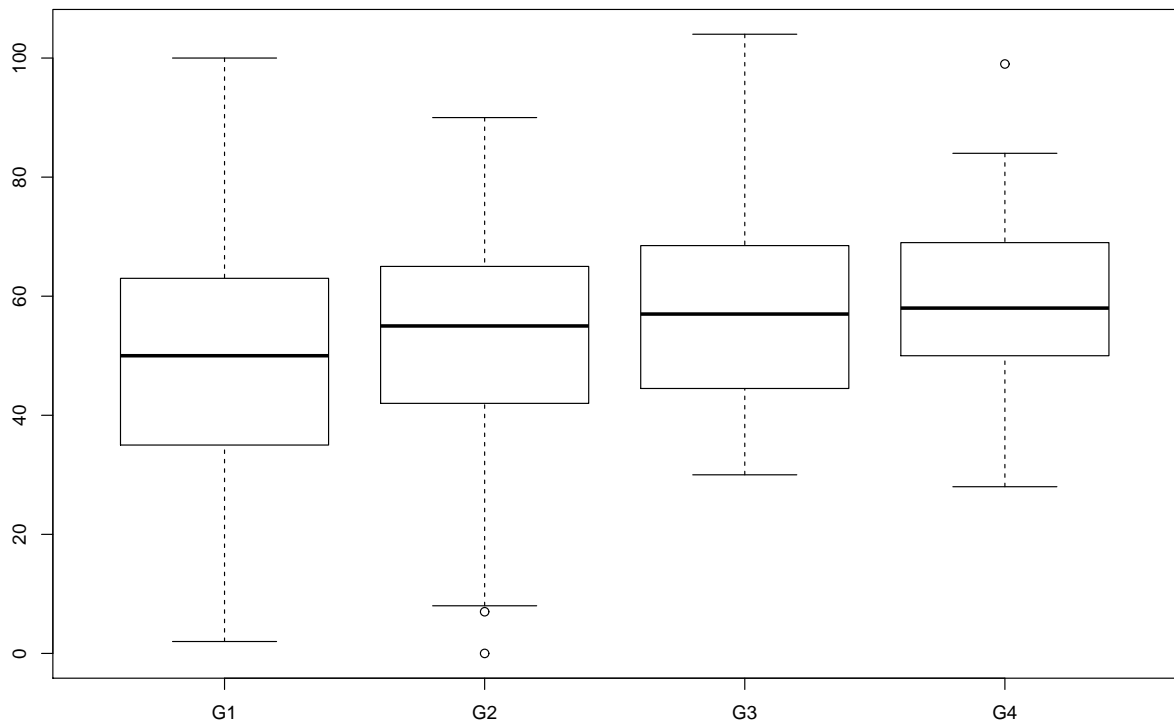}}
\caption{Boxplots for the four groups compared in the Well Elderly 2 study. G1=high school graduate, G2=some college or technical school, G3=4 years of college and G4=post-graduate school.}
\end{figure}
\section{Concluding Remarks}

All indications are that if the cardinality of the sample space is $N > 20$ and 
the sample size is $n \ge 30$, reasonably good control over the
Type I error probability will be achieved using method Q when the goal is to compare the population medians. 
When comparing the quartiles, now $n \ge 50$ might be required.
Of course, simulations do not guarantee that this will be the case for all practical 
situations that might be encountered. But the main
point is that no other method has been found that performs even tolerably well  in simulations when tied values are likely to occur, except for the
special case of $J=2$ groups. 

There are many variations of the
method used here. For example, 
in various situations, weighted bootstrap methods have 
been suggested when dealing with robust estimators;
see for example 
Salibian-Barrera and Zamar (2002) and the papers they cite. There are several alternatives
to the Harrell--Davis estimator that use a weighted sum of all the order statistics, and
there are variations of the projection distance that was used. 
perhaps there are situations where some combination of these methods provide  a practical
advantage over the method used here, but this remains to be determined. 
The main point is that  the method in the paper is the only known
method that continues to perform reasonably well when there are tied values.

Finally, method Q can be applied with the R function Qanova, which 
 has  been added to
 the Forge R package WRS.
This function is also 
stored in the file Rallfun-v28, which  can be downloaded from
  http://dornsife.usc.edu/cf/labs/wilcox/wilcox-faculty-display.cfm.

\begin{center}
REFERENCES
\end{center}


Bradley, J. V. (1978) Robustness? {\em British Journal of Mathematical and}
 {\em Statistical Psychology}, {\em 31}, 144--152.

Bonett, D. G. \& Price, R. M. (2002). Statistical inference for a linear  function of medians: Confidence intervals, hypothesis testing, and  sample size requirements. {\em Psychological Methods, 7}, 370--383.


Clark, F., Azen, S. P., Zemke, R., Jackson J., Carlson, M., Mandel, D., Hay, J., Josephson, K., Cherry, B., Hessel, C., Palmer, J., \& Lipson, L . (1997). Occupational therapy for independent-living older adults. A randomized controlled trial. {\em Journal of the American Medical Association,, 278}, 1321--1326.

Cliff, N. (1996). {\em Ordinal Methods for Behavioral Data Analysis}.  Mahwah, NJ: Erlbaum.

Dielman, T., Lowry, C. \& Pfaffenberger, R. (1994). A comparison of quantile estimators. {\em Communications in Statistics--Simulation and}
{\em Computation, 23}, 355--371.

Doksum, K. A. \& Sievers, G. L. (1976). Plotting with confidence: graphical comparisons of two populations. {\em Biometrika, 63}, 421--434.

Donoho, D. L. \& Gasko, M. (1992). Breakdown properties of the location estimates based on halfspace depth and projected outlyingness. {\em Annals of}
 {\em Statistics, 20}, 1803--1827.


Frigge, M., Hoaglin, D. C. \& Iglewicz, B. (1989). Some implementations of the  boxplot.  {\em American Statistician, 43}, 50--54.

Fung, K. Y., 1980. Small sample behaviour of some nonparametric  multi-sample location tests in the presence of dispersion differences.
{\em   Statistica Neerlandica, 34}, 189--196.

Harrell, F. E. \& Davis, C. E. (1982). A new distribution-free quantile estimator. {\em Biometrika, 69}, 635--640.

He, X., Simpson, D. G. \& Portnoy, S. L. (1990). Breakdown robustness of tests. {\em Journal of the American Statistical Association, 85}, 446--452.

Hettmansperger, T. P. (1984). {\em Statistical Inference Based on Ranks}.  New York: Wiley.

Hoaglin, D. C. (1985). Summarizing shape numerically: The g-and-h distribution. In D. Hoaglin, F. Mosteller \& J. Tukey (Eds.) {\em Exploring Data Tables} {\em Trends and Shapes}. New York: Wiley, pp. 461--515.

Hochberg, Y. (1988). A sharper Bonferroni procedure for multiple tests of 
 significance. {\em Biometrika, 75}, 800--802.

Jackson, J., Mandel, D., Blanchard, J., Carlson, M., Cherry, B., Azen, S., Chou, C.-P.,  Jordan-Marsh, M., Forman, T., White, B., Granger, D., Knight, B., \& Clark, F. (2009). Confronting challenges in intervention research with ethnically diverse older adults:the USC Well Elderly II trial. {\em Clinical Trials, 6}  90--101.

 
 
 
Liu, R. G. \& Singh, K. (1997). Notions of limiting P values based on data
 depth and bootstrap. {\em Journal of the American Statistical Association, 92},
 266--277.

Parrish, R. S. (1990). Comparison of quantile estimators in normal sampling.
 {\em Biometrics, 46}, 247--257.

Pratt, J. W. (1968). A normal approximation for binomial, F, beta, and 
 other common, related tail probabilities, I. {\em Journal of the}
{\em American Statistical Association}, {\em 63}, 1457--1483.

Racine, J. \& MacKinnon, J. G. (2007). Simulation-based tests that can use  any number of simulations. {\em Communications in Statistics--Simulation}
{\em and Computation,  36}, 357--365.

Radloff L. (1977). The CES-D scale: a self report depression scale for research in the general
population. {\em Applied Psychological Measurement, 1}, 385-401.

Romano, J. P. (1990). On the behavior of randomization tests without a group invariance assumption. {\em Journal of the American Statistical Association,}
 {\em 85}, 686--692.

Salibian-Barrera, M. \& Zamar, R. H. (2002). Bootstrapping robust estimates of regression. {\em Annals of Statistics, 30}, 556--582.

Sfakianakis, M. E. \& Verginis, D. G. (2006). A new family of nonparametric quantile estimators.
{\em Communications in Statistics--Simulation and Computation, 37}, 337--345.

Sheather, S. J. \& Marron, J. S. (1990). Kernel quantile estimators. {\em Journal}
 {\em of the American Statistical Association, 85}, 410--416.

 Wilcox, R. R. (2006). Comparing medians. {\em Computational Statistics} {\em \& Data Analysis, 51}, 1934--1943.

 Wilcox, R. R. (2012). {\em Introduction to Robust Estimation and Hypothesis Testing} (3rd Edition). San Diego, CA: Academic Press.

Wilcox, R. R.,  Erceg-Hurn, D., Clark, F. \& Carlson, M. (2014).  Comparing two  independent groups via the lower and upper quantiles. {\em Journal of Statistical}\\
 {\em  Computation and Simulation, 84}, 1543-1551.  DOI: 10.1080/00949655.2012.754026.

\end{document}